\documentclass[12pt]{article}
\usepackage[utf8]{inputenc}
\usepackage[authoryear]{natbib}

\usepackage[left=0.75in,right=0.75in,top=0.75in,bottom=0.75in]{geometry}
\usepackage{authblk} 
\usepackage{graphicx} 
\usepackage{comment}
\usepackage[bookmarks=false]{hyperref}
\usepackage{amsmath}
\usepackage{amssymb}
\usepackage[dvipsnames]{xcolor}
\usepackage{lineno}
\usepackage{tabularx}
\usepackage{soul}
\usepackage[font=scriptsize]{caption}

\title{Globally distributed environmental justice resistance against extractive corporate networks}

\author[1,2]{Dario Cottafava\footnote{Contact: dario.cottafava@uab.cat; jnicolas@unam.mx; marcelllaveropasquina@gmail.com}}
\affil[1]{\small Faculty of Economics and Business, Department of Business, Universitat Autonoma de Barcelona, Plaça Cívica, Bellaterra,
Barcelona, 08193, Spain.}
\affil[2]{\small Center for Studies and Research in Social Entrepreneurship and Innovation (CREIS), Plaça Cívica, Bellaterra,
Barcelona, 08193, Spain.}

\author[3]{José R. Nicolás-Carlock}
\affil[3]{\small Instituto de Física, Universidad Nacional Autónoma de México, Ciudad de México, 04510, México.}

\author[4,5]{Marcel Llavero-Pasquina}
\affil[4]{\small Institut de Ciència i Tecnologia Ambientals (ICTA–UAB), Universitat Autonoma de Barcelona, Edifici ICTA-ICP, Carrer de les Columnes s/n, Campus de la UAB, Cerdanyola del Vallès, 08193, Spain.}
\affil[5]{\small Departament de Biologia Evolutiva, Ecologia i Ciències Ambientals, Facultat de Biologia, Universitat de Barcelona,  Barcelona, 08028, Catalonia, Spain}

\author[4,6]{Joan Martínez Alier}
\affil[6]{\small Departament d'Economia i d'Història Econòmica, Facultat de Economia y Empresa, Universitat Autonoma de Barcelona, Plaça Cívica, Bellaterra,
Barcelona, 08193, Spain.}

\date{}

\begin{document}
\maketitle


\begin{abstract}
Is environmental injustice driven by a global network of multinational companies? Is there a global environmental justice movement resisting extractivism? This study addresses these questions through a social network analysis of firms and Environmental Justice Organizations (EJOs) involved in socioenvironmental conflicts. Using Environmental Justice Atlas (EJAtlas) data, we examine 3,396 conflicts involving 6,244 companies and 11,231 EJOs. The findings reveal evident structural asymmetries. Corporations form cohesive, sector-specific global networks that support coordination across regions. Conversely, EJOs form a larger, decentralized, and self-organized network characterised by sparse localised interconnections that enable regional coordination across sectors. A limited number of international EJOs involved in many conflicts worldwide weaves a global environmental justice network to match the global spread of multinational corporations and supply chains. This research study shows that these contrasting network structures reflect the distinct purposes of their constituent actors: sectoral expansion for corporations and environmental and territorial defense for EJOs.
\end{abstract}

\medskip
{\small Keywords: \textsl{EJAtlas, Multinational Companies, Extractivism, Environmental Defenders, Network Analysis, Complex Systems}}


The industrial economy is still growing. Since it is not circular but entropic \citep{martinez2023land}, even without further economic growth, new energy and material resources are extracted from ever-expanding ``commodity frontiers'' \citep{Bosma2024}, which generates yet more waste and emissions, regardless of the increased efficiency of extraction or production \citep{polimeni2015myth}. The addition of renewable energy to existing fossil fuel extraction \citep{fressoz2024sans}, increased demand for critical raw materials \citep{IEA2021,dowling2025mirages}, industrial agriculture \citep{busscher2020environmental}, infrastructural mega-projects \citep{stammler2016resources, orihuela2025megaprojects} and even waste management policies \citep{demaria2025contested} are accelerating the expansion of extractivism globally \citep{atapattu2018extractive}. 

In turn, this expansion generates unequal exchange of environmental costs and benefits between territories and populations \citep{temper2018perspective, hickel2022imperialist, llavero2025driving}. Such asymmetries, produced through new ``modes of extraction'' \citep{bunker1985underdeveloping}, generate constantly new ecological distribution conflicts \citep{martinez2021mapping}. Over the past decade, the Environmental Justice Atlas (EJAtlas) has become the most comprehensive global database documenting environmental conflicts  \citep{temper2015mapping,temper2018perspective,menton2020environmental}. Developed through a collaborative effort among researchers, activists, and environmental justice organizations (EJOs), the EJAtlas (\url{https://ejatlas.org/}) collects structured information on more than 4,500 environmental conflicts worldwide, including data on affected communities, corporations, governmental actors, and civil society organizations involved in each case \citep{Martinez-Alier03052016, temper2018perspective, martinez2021mapping, cantoni2025ejatlas}.

An underexplored dimension concerns the roles of companies and EJOs. Previous research shows that environmental conflicts are not merely technical disputes over resource management but are deeply embedded within broader structures of global political power and economic governance \citep{Carroll2003,vitali2014community,aydin2017network,scheidel2020environmental}. Uniquely, the EJAtlas allows researchers to study environmental conflicts not only as isolated events but as manifestations of systemic global dynamics \citep{llavero2025driving,cantoni2025franccafrique}. Indeed, environmental conflicts are inherently relational phenomena in which multiple actors interact, cooperate, or oppose each other \citep{temper2015mapping}. Despite the growing body of EJAtlas-based research, most studies focused on geographic distribution \citep{martinez2021mapping, llavero2025driving} or socio-ecological impacts of conflicts \citep{llavero2025false}, rather than on the structural relationships between the organizations participating in them. Only one study has previously examined the relational architecture of organizations involved in environmental conflicts for a subset of EJAtlas mining cases \citep{aydin2017network}. This leaves a gap regarding the structure and function of company and EJO networks across sectors globally. Therefore, originally motivated by the overarching question ``Is there a Global Environmental Justice (EJ) movement?'' \citep{Martinez-Alier03052016}, this research aims to answer: (A) How are EJOs involved in environmental conflicts networked at the global level?, and (B) What are the main structural differences between the global networks of EJOs and companies involved in environmental conflicts? 

To address these questions, network science models social systems as actors (nodes) connected through relationships (edges), revealing structural properties that cannot be observed from actors in isolation \citep{Newman2006, blondel2008fast, castellano2009statistical, fortunato2010community, barabasi2011bursts}. From this perspective, environmental conflicts can be understood as part of a broader network linking corporations, governments, investors, and civil society organizations across territories and sectors \citep{aydin2017network}, composing a complex \citep{ladyman2013complex, de2019complexity}, multilayer system   \citep{de2022multilayer, aleta2025multilayer} in which emergent structural properties arise \citep{de2019complexity}. Specifically, by analysing different centrality metrics \citep{castellano2009statistical} this study explores three EJAtlas-derived network representations: 1) actor networks linked by shared conflicts, 2) conflict networks linked by shared actors, and 3) actor networks aggregated by country or region 

Our study reveals pronounced asymmetries between the global networks of corporations and EJOs. First, 81\% of EJOs belong to a single connected network, compared with its counterpart that only connects 59\% of the companies. Second, the two networks exhibit contrasting sectoral structures: companies cluster strongly within specific sectors, such as mining and fossil fuels, whereas EJOs are more inter-sectoral. Third, geographical aggregation shows that corporate interactions are concentrated within the Global North and along North–South relations, while EJOs interactions are predominantly localized in the Global South. A small number of international EJOs (IEJOs) involved in hundreds of conflicts act as bridges connecting geographically distant struggles. Finally, we argue that these contrasting network architectures emerge from the distinct purposes of their constituent actors: sectoral expansion for corporations and territorial defense for environmental defenders.

These findings have important implications. They suggest that EJ should be understood not only as a collection of isolated local struggles, but also as a collective phenomenon embedded in global networks of actors and power relations \citep{helbing2013globally}. Accordingly, environmental conflicts cannot be fully understood without considering how involved actors are positioned within broader relational structures shaped by the distinct objectives and aims \citep{aydin2017network}.

\begin{figure}[ht!]
\centering
\includegraphics[width=1.0\textwidth]{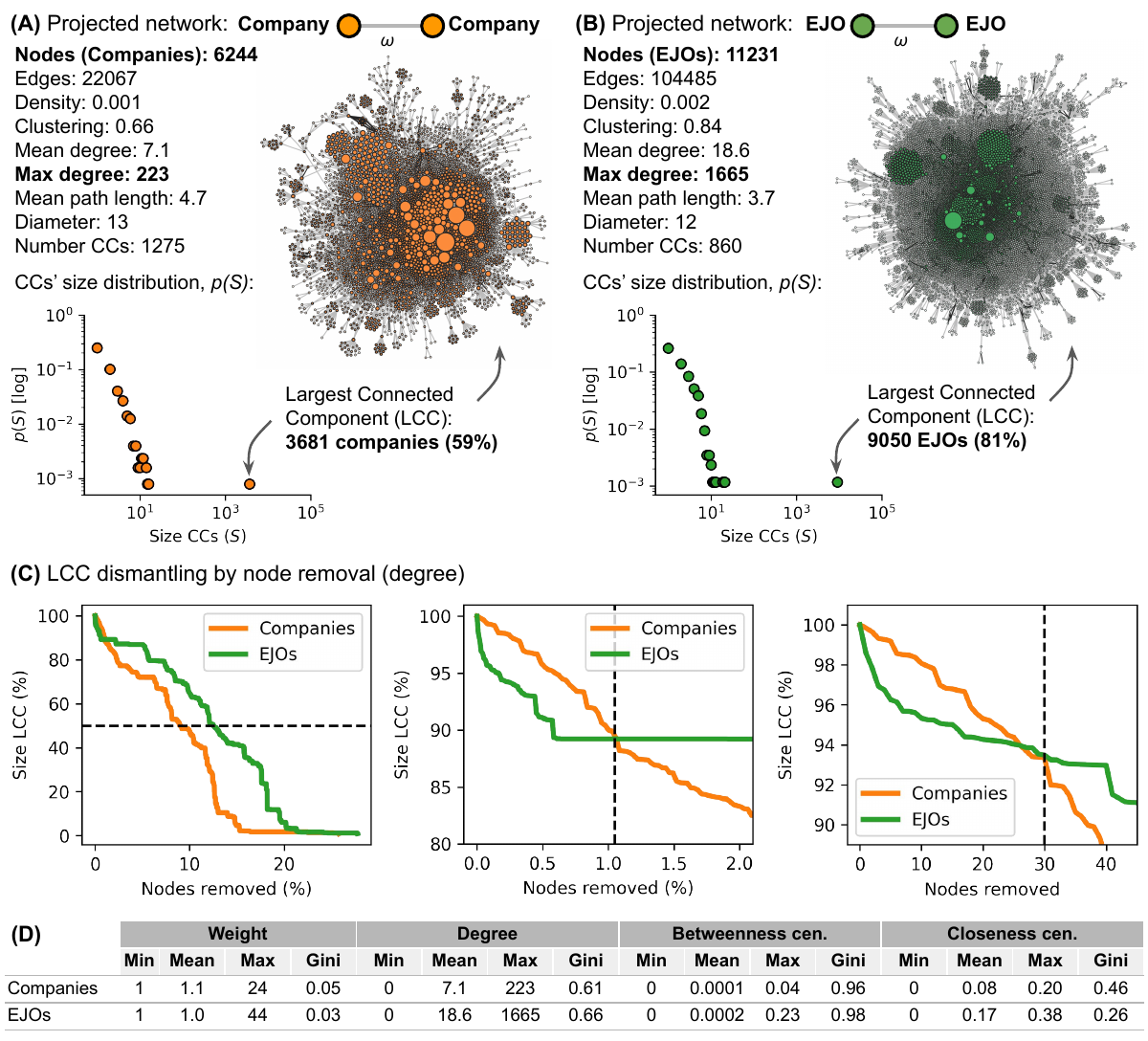}
\caption{\textbf{Company and EJO projected networks.} In the (A) companies and (B) EJOs projected networks, edges represent co-occurrence in a conflict for two nodes (either two companies or EJOs) while the weight measures the $\#$ of shared conflicts among two actors. Both in (A) and (B), the size of each node is proportional to their degree (note that the scales are different for each network).
For each projected network, several metrics are shown: $\#$ of edges and nodes, 
density (i.e., the  proportion of existing edges in a network compared to the total number of possible edges), clustering (the degree to which nodes in a graph tend to cluster together), mean and max degree (average and maximum number of links), average mean path length (the average $\#$ of steps between two nodes), diameter (the longest of all shortest paths between any two nodes in a graph), and number of Connected Components (CCs) and the Largest Connected Component (LCC) (the single group that contains the greatest number of interconnected nodes). The scatter plots below the two networks shows the CCs size distribution $p(S)$ vs the size of the CCs $S$ for both networks. (C) Robustness test for the LCC by removing one-by-one the top-degree nodes. The first two plots show the LCC size (\% of the original size) vs the percentage of nodes removed, while the third one represents the absolute number of nodes removed. (D) Summary statistic data, i.e., weight ($\#$ of shared conflicts among two actors), degree ($\#$ of edges), betweenness ($\#$ of shortest paths between any couple of nodes in the network that pass from a specific node) and closeness (reciprocal of the sum of the shortest path distances between a target node and every other node in the network) centrality measures. Each centrality reveals a different property: weight and degree reveal how much a node is connected to others nodes locally, betweenness reveal which nodes act as gatekeeper, while closeness shows the average distance among nodes and how compact is the whole network.}
\label{fig 5: actor-actor network}
\end{figure}

\subsubsection*{Company and EJO Networks: the emergence of a resilient global EJ movement}\label{ref: monolayer neworks}

The EJAtlas database provides us with 3,396 socio-environmental conflicts involving 6,244 companies and 11,231 EJOs worldwide (see The EJAtlas Database and Data Curation in Methods). We use conflicts to construct actor co-occurrence networks by projecting event–actor bipartite networks, in which companies or EJOs are connected to the conflicts in which they participate (see Data Analysis in Methods). 

The projected company (Fig.~\ref{fig 5: actor-actor network}A) and EJO (Fig.~\ref{fig 5: actor-actor network}B) co-occurrence networks exhibit important structural characteristics. Both networks exhibit fat-tail degree distributions, indicating the existence of relevant outliers or central hubs, i.e., companies and EJOs with very high degree (see also the degree distributions in Supplementary Fig. B1A-B1B). Furthermore, both networks are conformed by multiple connected components (or subnetworks) indicating some fragmentation in participation across global conflicts. Nevertheless, their largest connected components (LCCs) reveal the existence of a global environmental justice network: the EJO LCC contains 81\% of all EJOs, compared with 59\% of companies in the company LCC. The EJO LCC also connects actors more efficiently. Its diameter and mean path length are 12 and 3.7, respectively, compared with 13 and 4.7 for the company LCC. This is notable because EJOs participate, on average, in fewer conflicts than companies (1.59 versus 1.89 conflicts per actor) yet a much larger share of them belongs to the LCC (Supplementary Fig. B1C). The higher mean degree of EJOs (18.6 versus 7.1) and network density (0.002 versus 0.001) further support this result.

Other centrality measures reinforce these differences (Fig.~\ref{fig 5: actor-actor network}D). EJOs have a slightly lower mean edge weight than companies (1.0 versus 1.1), but a higher maximum weight (44 versus 24), indicating that most actors co-participate in few conflicts, while a small number of EJOs repeatedly appear together. Edge-weight inequality is low and comparable in both networks, with Gini coefficients of 0.05 for companies and 0.03 for EJOs. Degree is substantially higher in the EJO network, both on average (18.6 versus 7.1) and at the maximum (1,665 versus 223). Its higher degree Gini coefficient (0.66 versus 0.61) confirms that a small number of EJOs are exceptionally well connected. Three EJOs are clear outliers, acting as bridges among otherwise distant parts of the network (Supplementary Fig. B1B). EJOs also have higher closeness centrality (mean 0.17, maximum 0.38) than companies (mean 0.08, maximum 0.20) and a lower closeness Gini coefficient (0.26 versus 0.46), indicating that they are generally closer and more evenly positioned within the network. Together, these results suggest a globally distributed and relatively decentralized environmental justice movement connected by a few important bridging organizations.

To assess how the LCCs depend on high-degree nodes, we performed a targeted robustness test by sequentially removing nodes in descending order of degree (Fig.~\ref{fig 5: actor-actor network}C). The EJO LCC remains more resilient than the company LCC after the removal of the 30 highest-degree nodes. However, during the initial removal of the very top hubs (less than 1\% of all nodes) the EJO LCC declines more sharply. Thus, although the EJO network is more robust overall, its initial connectivity depends more strongly on a small set of central hubs.

\begin{figure}[ht!]
\centering
\includegraphics[width=1.0\textwidth]{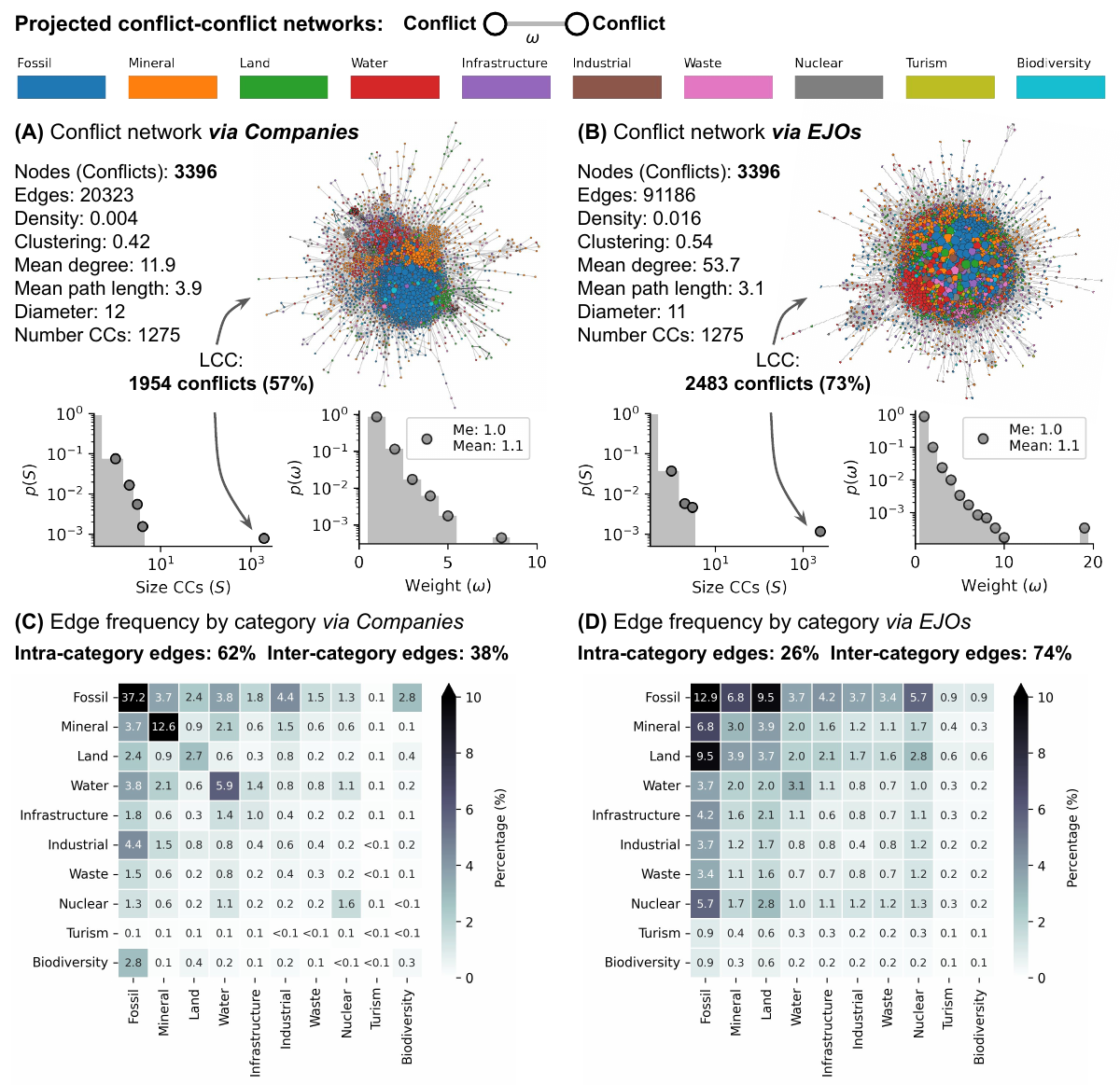}
\caption{\textbf{Projected conflict networks (\textit{via companies} and \textit{via EJOs}) and intra- and inter-category edge frequencies.} (A) Conflict projected network \textit{via companies}. (B) Conflict projected network \textit{via EJOs}. In these projected networks (Supplementary Fig.~\ref{fig 1: flowchart} and \ref{fig 2: visual representation} for the methodological details and logic), edges represent (if any) shared organizations while the weight counts the $\#$ of shared organization between two different conflicts. Colors represent different conflict categories. The Graphs below represent the CC's size and weight distribution, specifically highlighting the LCC's size: 57\% (1954 conflicts) and 73\% (2483 conflicts) for the projected network \textit{via Companies} and \textit{EJOs} respectively.
(C) Edge frequency by category \textit{via Companies}. (D) Edge frequency by category \textit{via EJOs}. In both matrices, intra-category and inter-category are reported. Intra-category edge frequency (elements on the diagonal) measures the frequency of edges among conflicts within the same category, while the inter-category edge frequency (elements out of the diagonal of the matrix) measures the percentage of edges between conflicts with two different categories. Globally, the intra-category edge frequency (sum over the diagonal) in the conflict network \textit{via companies} is higher than in the network \textit{via EJOs} (62\% vs 26\%), while the inter-category edge frequency (sum of cell out of the diagonal) is lower (38\% vs 74\%). 
}
\label{fig 6: conflict-conflict network}
\end{figure}

\subsubsection*{Conflict networks: company-related conflicts clustered by category}\label{sec: conflict-conflict category networks}

In this analysis, companies and EJOs are used as bridges to create conflict co-occurrence networks by projecting the previously defined event-actor bipartite networks. This produces two networks (Fig.~\ref{fig 6: conflict-conflict network}A--B): conflicts connected \textit{via companies} and conflicts connected \textit{via EJOs} (see Data Analysis in Methods). 

In the network of conflict \textit{via EJOs}, conflicts show no clear clustering, mixing among categories, while in the network \textit{via companies}, firms cluster based on their category. The network \textit{via EJOs} counts approx 91,000 edges, a density of 0.016 and a mean degree of 53.7. By comparison, the network \textit{via companies} counts approx 20,000 edges with a density of 0.004 and mean degree of 11.9. The largest connected component in the EJO network, with 2,483 conflicts (approximately 73\% of the total), is larger compared with the company-mediated network which counts 1,954 conflicts (57\% of the total).

The frequencies of intra-category and inter-category edges further reveal both commonalities and differences (Fig.~\ref{fig 6: conflict-conflict network}C--D). In both networks, the highest edge frequencies involve the Fossil, Mineral, Land, and Water categories, whereas Tourism and Biodiversity are nearly negligible, partly reflecting the composition of the database. The main difference lies in category mixing. In the conflict network \textit{via EJOs} contains more inter-category than intra-category connections, indicating that EJOs tend to link conflicts across environmental sectors. In contrast, in the conflict network \textit{via companies}, intra-category edges dominate (for example, 27\% within Fossil, 9.1\% within Mineral, and 4.3\% within Water) while cross-category frequencies are much lower (such as 2.7\% between Fossil and Mineral and 1.8\% between Fossil and Land). This produces cohesive sectoral clusters. 

Thus, EJOs create a broad and heterogeneous cross-sectoral network with an inter-category edge frequency of 74\%, whereas companies connect conflicts primarily within the same category with an intra-category edge frequency of approximately 62\%.

\subsubsection*{Aggregated networks by country: EJOs clustered by country}

\begin{figure}[ht!]
\centering
\includegraphics[width=1.0\textwidth]{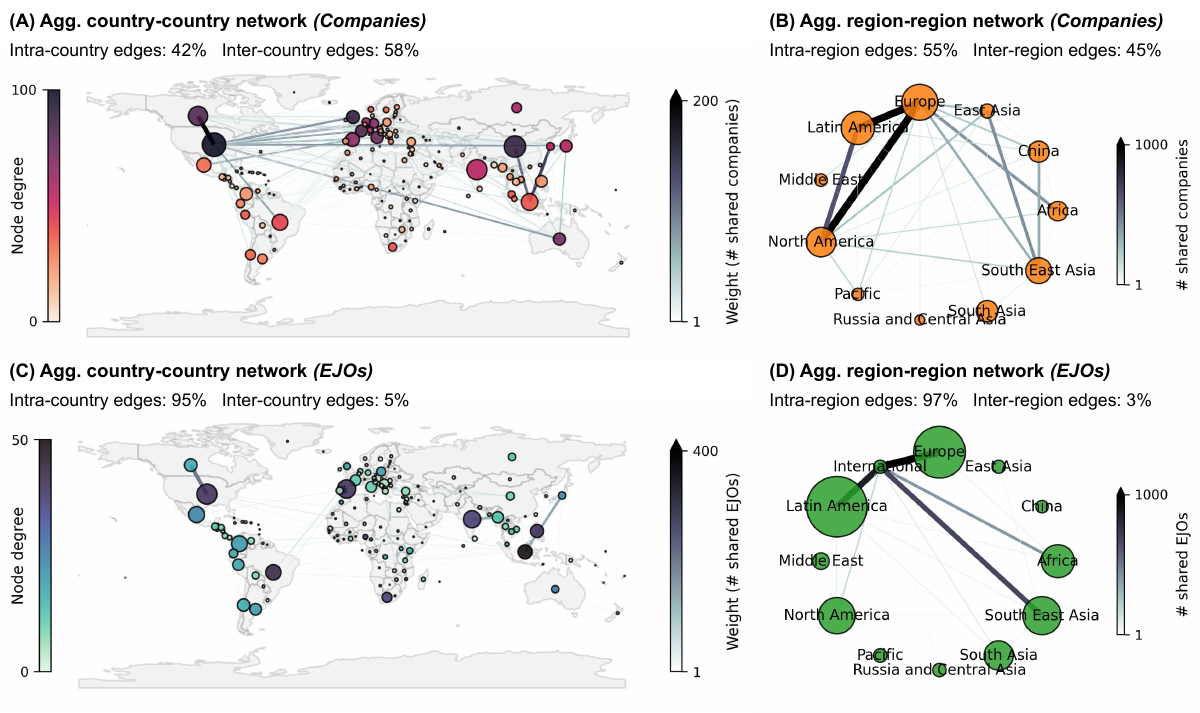}
\caption{
\textbf{Company and EJO projected network aggregated by country or region.} (A) Company projected networks aggregated by country. Country/region node size represents the \# of actors headquartered in a specific country, while node color - from darker to lighter - evidences the country degree (the $\#$ of distinct countries connected to a given country via EJO or company co-occurrence in a conflict). This representation allows to decouple how many actors are present within a country from the geographical extent of their interactions. Edge weight is shown from thinner lighter to thicker darker lines, highlighting stronger links among countries such as the strong ties among USA and Canada, Brazil or many European countries in the company networks. (B) Company projected networks aggregated by region. A clear North-South dependency triangle with strong ties among North America, Europe and Latin America emerge. (C) EJO projected networks aggregated by country. No strong ties among countries emerge from the EJO projected network except a few exception in Asia among Japan and Indonesia. IEJOs are not represented in the country map for graphical reason. (D) EJO projected networks aggregated by region. The centrality of IEJOs in linking different regions with very strong ties mainly towards Latin America, Europe and South East Asia is evident. 
In every graph intra-country/region (within the same country/region), and inter-country/region edges are reported above the graph: companies connect more with companies from other countries (58\% inter-country vs 42\% intra-country edges) while, EJOs link with other EJOs within the same country (95\% intra-country vs 5\% of inter-country edges).}
\label{fig 8: country region aggregation}
\end{figure}

In this section, company and EJO networks are aggregated by country/region based on the headquarter of the organization while edges represent the edges between two organizations, either EJOs or companies, via a co-occurrence in a conflict. In all visualization in this section, any information regarding the conflicts location is lost (see Data Analysis in Methods). 

Most companies are located in the Global North, particularly in USA, Mexico, Canada, UK, Spain, France, Italy and China. Several countries in the Global South, including India, Indonesia, Brazil, and Colombia, also host a substantial number of companies with strong ties to the Global North (Fig.~\ref{fig 8: country region aggregation}A). Notably, decoupling the \textit{country degree} (node color) from the number of companies (node size), reveals the well-known North-South dependency \citep{ghosh2019dependency, cantoni2025franccafrique}. Most Global North countries (particularly USA, Canada, UK, France, Japan), along with China, have high country degree. By contrast, Global South countries have few companies and a low degree, with some exceptions. For example, Brazil, India and Indonesia host many companies but this is not reflected in their degree. This indicates that companies from the Global North have connections with more foreign companies than companies from the Global South. More specifically, Global North multinationals frequently co-occur in environmental conflicts, most of which take place in the Global South \citep{llavero2025driving}. Although the EJAtlas does not systematically document direct collaboration among such firms, our findings suggest that there is a cohesive network of Global North extractive multinational companies woven together by shared interests and systemic economic functions. 

In contrast, Fig.~\ref{fig 8: country region aggregation}C shows that EJOs tend to operate locally within their home countries with less international connections. This is reflected in less ties among countries and lower weights compared to companies. The frequencies of intra-country edges (within the same country) and inter-country edges (between different countries) illustrate this difference: inter-country edges account for only 5\% in the EJO network, while approximately 58\% in the company network.

\subsubsection*{Aggregated networks by regions: international EJOs strengthen the global EJ network}

At the regional level, EJOs are concentrated in Latin America, Europe, North America, followed by South East Asia, South Asia and Africa. A further structural difference emerges at this scale. The international region contains relatively few EJOs but shows strong connections with Europe, Southeast Asia, and Latin America forming a T-shaped pattern, while ties with other regions remain weak. By contrast, the company network shows the strongest links among Europe, North America and Latin America (forming a triangular shape), and between East Asia and Southeast Asia. These configurations suggest exploitative relationships, from East Asian economies towards South East Asia, and similarly from Europe and North America towards Latin America. Both in Latin America and South East Asia, we see multinational extractive companies co-occurring in the same conflicts with small local companies, or influential domestic corporate groups. To a much lesser extent, this architecture is also present between European multinationals and African companies, but not between Chinese and African companies. There is relatively little co-occurrence between Chinese and Global North companies. Finally, inter-regional edges account for only the 3\% in the EJO network compared with the 45\% of all connections in the company network.

\section*{Discussion}\label{sec: discussion}

\subsubsection*{Local struggles are interconnected through shared narratives: an emergent globality}

Is there a global EJ movement? The question of whether a global EJ movement exists has been central to EJ literature since its inception \citep{Martinez-Alier03052016}. Our findings provide a nuanced answer supporting the existence of a robust, resilient and emergent global EJ movement. From a networks perspective, the EJO network contains a LCC encompassing more than 80\% of all the EJOs, despite EJOs participating in a small number of conflicts in average (Fig.~\ref{fig 5: actor-actor network}). These results reveal two contradictory dynamics: EJOs are predominantly embedded in localized clusters, with intra-country edges representing 95\% of all links, reflecting the place-based nature of environmental conflicts \citep{sebastien2019resistance,Temper_2020}; while a limited number of IEJOs act as bridges connecting otherwise disconnected regions (see T-shape in Fig.~\ref{fig 8: country region aggregation}D). However, IEJOs are not the only drivers of global connectivity (Fig. \ref{fig 5: actor-actor network}C). Local EJOs contribute to the global structure through shared participation in conflicts and the horizontal diffusion of knowledge and narratives across cases. In this sense, the global EJ movement forms a globally-connected yet decentralized system, rather than a coordinated structure or a highly clustered core as that observed in the company network (Fig. \ref{fig 5: actor-actor network}). 

The self-organized structure of EJOs is consistent with theories of complex adaptive systems, in which global patterns emerge from decentralized interactions without hierarchical coordination \citep{ladyman2013complex, de2019complexity}. IEJOs act as important accelerators or amplifiers of connectivity, but the movement remains fundamentally distributed and rooted in local struggles. This perspective aligns with ``a subaltern, thick, embedded, and rooted cosmopolitanism [...] that combines the aspiration to achieve transnational and global justice with attentiveness to local struggles and realities as they actually exist'' \cite[p.435]{della2021transnational}. IEJOs are themselves an emergent feature of the global EJ movement. They occupy a critical niche that responds to the needs of local struggles to connect globally along a shared resistance to environmental injustice driven by globalization \citep{almeida2018globalization}. Thus, the niche matters more than the actor occupying it. The network is resilient because even if one of the IEJOs were to disappear, another actor would take that niche. The globally-connected yet decentralized structure is the product of territorially embedded struggles that share the purpose of resisting global drivers of environmental injustice (Faccion \textit{et al.}, forthcoming).

This results in a form of emergent \textit{globality}, where the overall coherence of the network does not depend exclusively on centralized or international EJOs, but arises from the accumulation of local interactions and relational ties. This structure is supported by the high number of inter-category edges in the EJO network and the robust and resilient LCC structure of the EJO networks (Fig. C1). Such a configuration is consistent with the conceptualization that local struggles are globally interconnected through shared narratives and common struggles \citep{lindell2009glocal, vanhaute2014globalizing}. Future case studies should identify shared political agendas, \textit{a common identity}, or a common set of beliefs, values and norms across global EJ movements \citep{Kimberly2012, della2021transnational} that mechanistically explain the emergent properties of the global EJ movement. Ironically, some authors have argued that the very process of globalization have provided the digital technologies and cultural homogeneity conditions for international movements to emerge \citep{almeida2018globalization,smith2013transnational}.

\subsubsection*{Place-based distributed resistance against multinational sectorial extractivism}

While a global EJ movement emerges structurally, it exhibits a fundamentally different network organization compared to corporate actors. The company network displays a smaller LCC with a lower density and a higher diameter than the EJOs network; however the core of the network structure appear more clustered with multiple actors exhibiting high degree and forming tightly clustered communities (Fig. \ref{fig 5: actor-actor network}). This reflects the systemic nature of global extraction and production networks and transnational corporate structures \citep{vitali2014community,Carroll2003}, where firms operate across multiple regions, often participating in several conflicts simultaneously. Such structural cohesion, further supported by the network by category (Fig. C1), suggests a strategic alignment among corporate actors with a risk to reproduce harmful practices across different countries. Further research is needed to assess the systemic risks of reproducing harmful corporate activities, either in depth through case studies \citep{mliless2022reporting} or sectorial analysis \citep{avila2018environmental}, or via structured risk and corruption indicators \citep{Carlock2021,nicolas2021conspiracy, luna2020corruption,christiansen2025corruption,williams2024transnational}. 

Geographically, the company network shows more international connections, specially between Global North companies, and between these and their partners in the Global South. In contrast, the EJO network shows more international connections between Global South organizations, and with IEJOs (Fig.~\ref{fig 8: country region aggregation}). The comparison between the two networks highlights a clear asymmetry of structural power. These global structures may reflect broader dynamics of ecological distribution conflicts, ecologically unequal exchange and global North–South dependencies \citep{bunker1985underdeveloping,martinez2021mapping,hickel2022imperialist,llavero2025driving, cantoni2025franccafrique}, where economic power is concentrated in transnational corporate networks headquartered in the North. Globally, corporations tend to cluster according to their category presenting a higher intra-category edge frequency (Fig. C1) while EJOs cluster based on their location showing a higher intra-country edge frequency (Fig.~\ref{fig 8: country region aggregation}).

\subsubsection*{Divergent organizational logics lead to different global network structures}

\begin{table}[ht!]
\centering
\footnotesize
\setlength{\tabcolsep}{4pt}
\renewcommand{\arraystretch}{1.2}
\resizebox{\textwidth}{!}{
\begin{tabular}{p{3cm} p{3.2cm} p{5.5cm} p{3.2cm} p{5.5cm}}
\hline
\textbf{Concept} & \textbf{Corporations} & \textbf{Description} & \textbf{EJOs} & \textbf{Description} \\
\hline

Nature of the actor 
& Expansion logic 
& Oriented toward growth, capital accumulation, and infrastructure development across sectors and regions, often driven by global market  investment strategies. 
& Protection logic 
& Oriented toward defending territories, ecosystems, and communities from environmental harm, grounded in local needs and socio-ecological values. \\

Attachment rule 
& Sector-based clustering 
& Preferential attachment driven by sectoral specialization (e.g., mining, energy), leading to dense intra-sector connections and repeated co-occurrence in similar types of conflicts. 
& Place-based emergence 
& Attachment driven by territorial proximity and shared local conflicts, resulting in context-specific and spatially embedded connections. \\

Dynamics 
& Coordinated and strategic 
& Proactive planning and coordination across projects and geographies, supported by financial, institutional, and corporate governance structures. 
& Reactive and adaptive 
& Mobilization triggered by environmental conflicts, adapting dynamically to local conditions with limited prior coordination. \\

Aggregation level 
& Global integration 
& Embedded in transnational extraction and production and investment networks, enabling coordination and expansion across multiple regions and scales. 
& Local response with emergent globality 
& Primarily local in action, but globally connected through shared and commmon values. \\

Temporal logic 
& Continuous expansion 
& Long-term investment cycles and persistent involvement across projects, leading to sustained presence in multiple conflict contexts. 
& Event-driven emergence 
& Episodic activation in response to specific environmental conflicts, with varying duration depending on local conditions. \\

Network role 
& Systemic hubs and clusters 
& Formation of dense clusters and highly interconnected hubs, increasing systemic influence and the potential propagation of risk across the network. 
& Distributed connectors 
& Presence of a few bridging actors alongside many peripheral nodes, enabling emergent global connectivity despite fragmentation. \\

\hline
\end{tabular}
}
\caption{Divergent organizational logics of corporations and EJO networks: conceptual dimensions for systemic risk future studies.}
\label{tab:actor_logic}
\end{table}

These structural differences can be interpreted in light of EJOs and companies fundamentally different organizational logics and objectives. Corporations primarily seek to maximize the expansion of infrastructure, extraction and production systems within specific sectors, such as mining, energy, or transport \citep{llavero2024political}. This sectoral specialization is reflected in the tendency for companies to co-occur within similar categories of conflicts, indicating coordinated or parallel involvement in projects characterized by comparable technological and economic logics. Such a structure suggests an intentional strategy toward increasing global integration, where corporate actors expand their activities across regions while maintaining sectoral coherence, reinforcing patterns of coordination and accumulation at global scale. These coordination and expansion processes towards accumulation remind us, drawing on \citet{georgescu-roegen1971entropy}, that economic processes are entropic: they transform low-entropy resources into high-entropy waste. For Georgescu-Roegen, this entropic degradation undermines the very conditions for sustained economic and social life. As explained by \citet{martinezzijin} extractivist transnational companies, as Zijin a growing metal Chinese company \citep{martinezzijin} or the giant Glencore \citep{neyra2026glencore}, act as entropy export-import actors, by free-riding on social and environmental unaccounted costs, generating ``environmental liabilities'' \citep{neyra2026glencore}, and by exporting low-entropy energy, materials but also human labour-time from the periphery to the core. These unaccounted costs are even more evident in the waste management sector. Emphasized by the ``recycling paradox'' as described by \citet{demaria2025contested} who analyzed over 70 conflicts worldwide, modern waste and circular economy policies are enhancing processes of \textit{capital accumulation by dispossession}, excluding informal waste pickers, and \textit{by contamination}, shifting unaccounted costs onto marginalized communities \citep{d2024accumulation}. 

In contrast, EJOs operate according to a fundamentally different rationale. Their primary objective is not expansion, but the protection of specific territories, and communities affected by socioecological harms. As a result, their emergence is often reactive and territorially-grounded, triggered by localized instances of environmental injustice. This gives rise to a network structure that resembles a \textit{distributed response system}, where EJOs appear and mobilize in specific locations without central coordination. As such, environmental conflicts are not only the expression of the expansion of the extraction frontier clashing against socio-ecological values. Environmental conflicts are also the expression of a collision of scales between the globalizing logic of the increasingly entropic industrial economy \citep{FischerKowalski01011997,Martinez-Alier02112022} and the place-based ancestral and traditional livelihoods linked to communities and territories. In this sense, EJOs can be metaphorically understood as \textit{white blood cells of socio-ecological systems} (quoting \citet{hanavcek2024we} ``We are protectors, not protestors''), acting as decentralized guardians that respond to threats as they arise. This analogy highlights both the strengths and limitations of the movement: while it enables rapid and context-sensitive responses, it also contributes to fragmentation and limits the capacity for coordinated action at global scale. 

Future studies should investigate other available information in EJAtlas and the underlying hypotheses in this article. Specifically, the structural differences summarized in Table~\ref{tab:actor_logic}, highlight divergent organizational logics that are only partially captured by this study and the current network representation. The coexistence of these two logics (expansive and coordinated for corporations, reactive and distributed for EJOs) ultimately shapes the asymmetric structure of the global environmental conflict network. Concluding, this study affirmatively answers the EJAtlas's mother question: a global EJ movement exists.


\pagebreak
\newpage

\section*{Methods}\label{sec: methodology}

\subsection*{The EJAtlas Database}

The EJAtlas, the most comprehensive global environmental conflict database, has been developed through a collaborative process between researchers, activists, and environmental justice organizations, and it systematically collects, via a crowdsourcing process, structured information on environmental conflicts worldwide including data on conflict characteristics, sectors, geographical location, and the actors involved \citep{temper2015mapping,Martinez-Alier03052016,temper2018perspective, martinez2021mapping, cantoni2025ejatlas}. The EJAtlas is specifically well-suited to address and analyse global structural differences of two key categories of actors: companies and environmental justice organizations. Companies frequently appear as project developers, investors, or operators in environmentally contested activities, particularly in sectors such as mining, fossil fuels, and large-scale infrastructure. Data on companies is systematically collected, and includes their country of origin. However, although we can certainly link a company to a conflict, we can not systematically differentiate the role a company has played in that conflict, whether it was the main operator, a contractor, a shareholder in the past, etc. Each of the EJAtlas cases also collects the names (and websites) of the involved environmental justice organizations, which can be generally defined as self-organised groups including local NGOs, grassroots organizations, and transnational advocacy groups that mobilize to defend environmental rights, local livelihoods, and ecological integrity \citep{martinez2021mapping, temper2015mapping, temper2018perspective, Martinez-Alier03052016}. 

For this study, we use the EJAtlas V3 dataset version from the 14th of July 2024, which originally contains information on 4,334 environmental conflicts. Our working dataset was obtained by selecting only conflicts with at least one EJO AND one company involved. Details about the preliminary cleaning and filtering process are reported in Appendix A in the Supplementary Information. As shown in Fig.~\ref{fig 3: ejatlas summary}A, the working dataset used in this study includes 3,396 conflicts across 164 countries, involving 6,244 companies and 11,231 EJOs, and covers a wide range of sectors (represented in different colors) such as mining, fossil fuels, infrastructure, waste management, and biodiversity conservation, reflecting the diversity of ecological distribution conflicts globally \cite{Martinez-Alier03052016,temper2018perspective,menton2020environmental}. Fig.~\ref{fig 3: ejatlas summary}B represents the time evolution of conflicts and their distribution worldwide (emphasizing the \# of conflicts per country). The EJAtlas mostly covers the first two decades of the 2000s, with a peak around 2010. At geographical level, Europe, North America and South America are overrepresented while most parts of Asia and Africa are underrepresented. These numbers are a direct consequence of the nature of the EJAtlas, since it derives from a participatory crowdmapping process that engages activists and researchers around the world. Finally, Fig.~\ref{fig 3: ejatlas summary}C shows the heatmap of the $\#$ conflicts per category and per geographical area, showing how the most represented categories are Fossil Fuels and Climate Justice, Mineral Ores and Building Materials Extraction, Biomass and Land Conflicts, and Water Management.

\begin{figure}[ht!]
\centering
\includegraphics[width=1.0\textwidth]{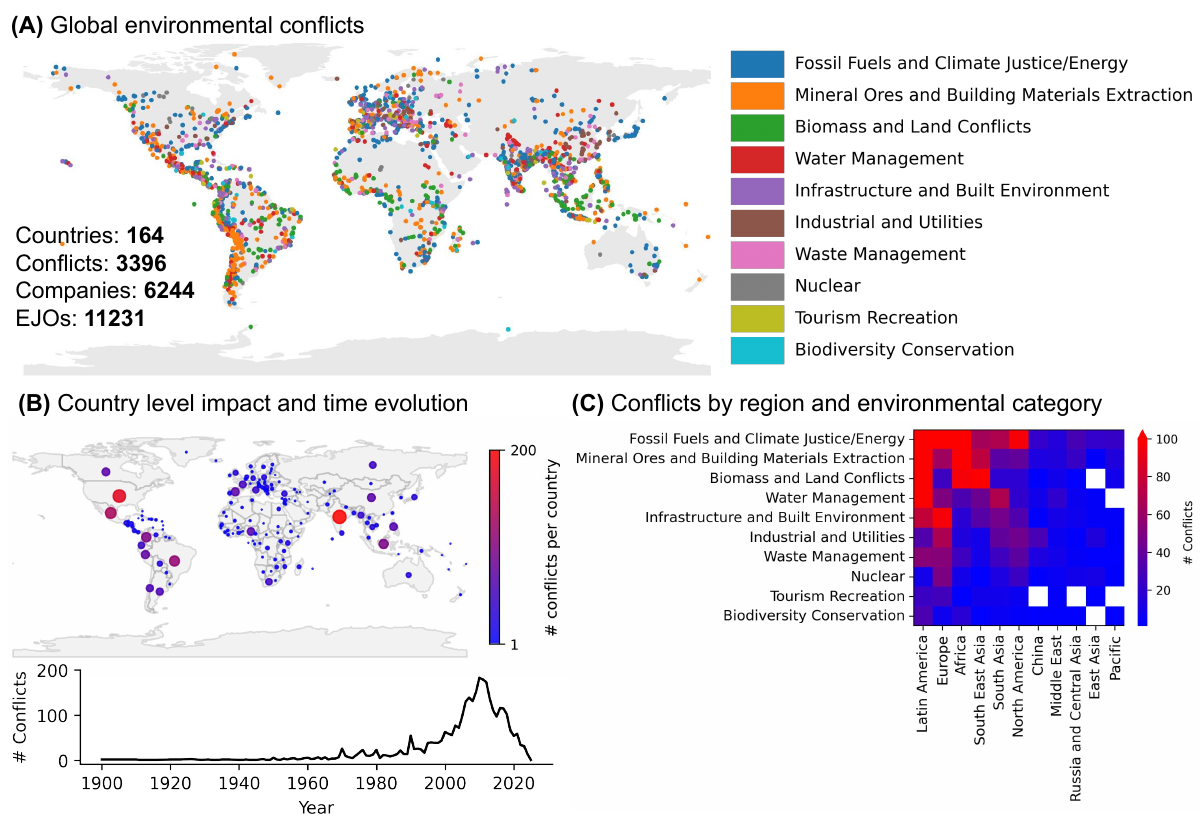}
\caption{\textbf{EJAtlas database summary.} (A) Geographical distributions of the environmental conflicts per category and summarizes the total number of countries, conflicts, companies and EJOs. After the initial cleaning process (see Appendix A in the Supplementary Information) the working dataset (only conflicts with at least one company and one EJO have been selected) includes 3,396 conflicts, 6,244 companies and 11,231 EJOs. (B) Time evolution of conflicts and its distribution worldwide (emphasizing the \# of conflicts per country). Conflicts are mainly located in North/South America and in South East Asia. (C) Heatmap of the \# of conflicts per category and per geographical area. The majority of conflicts are related to Fossil Fuels, Mineral Ores, Biomass and Water Management.}
\label{fig 3: ejatlas summary}
\end{figure}

\subsection*{Data Curation}
The company and EJO data in the EJAtlas have not been compiled into a comprehensive database before. This is not a trivial procedure as the crowdsourced data collection methodology of the EJAtlas leads to duplications and alternative spellings of company and EJOs names. The company dataset was recently curated by \cite{llavero2025driving} using a bespoke name matching algorithm to eliminate duplicates. Specifically for the purposes of this article, the algorithm minimises the number of false positive (merging two unrelated companies) hits and there was a manual validation of all company matches resulting in a company with more than one case to further minimise false negatives. 

A similar approach has been taken to produce, for the first time in this study, a global dataset of environmental justice organisations involved in environmental conflicts. The curation was complicated because the EJOs field is not structured and systematised as the companies field in the EJAtlas. But it was made easier because unlike companies EJOs do not have (as frequent) name changes, or mergers and acquisitions, or subsidiaries with different names. The EJOs name-matching algorithm, provided in the data repository, was coded in R and run in R version 4.5.2. The first step of the process is to separate the different EJOs named in each conflict detecting the specific delimiter in each case. The next step is to strip and assign the website for each EJO. The code also tries to eliminate any explanatory text that accompany a minority of EJOs, if the explanatory text is undistinguishable from the EJO then an apparently very long name will remain, and these are eliminated from the dataset (with a cutoff between 3 and 100 characters for EJO names). The next step is the name matching to consolidate under a unique ID the EJOs that appear in different cases. This is the critical step, specifically for the purpose of this study. This step follows the same logic of the companies name-matching algorithm \citep{llavero2025driving}, minimising false positives by creating a query for each entry that removed the common words present in the sample (ie. Organisation, Association, Green, etc), and only linking entries that part of their string fully matched the specific query of another entry. Please note that our name-matching methodology would consolidate "Ecoguardians Italy" with "Ecoguardians Spain" under the same unique EJO. There was a manual validation of all resulting consolidated EJO entries participating in more than 4 environmental conflicts to minimise false positives. Minimising false positives makes the name matching less stringent, and therefore leads to more false negatives (ie. two EJOs that should be linked, but were not), but this is a more conservative position than a more stringent alternative. After validation, all consolidated EJO entries were assigned a unique id, and a list of conflicts they were involved in. The EJOs database was anonymised for security and privacy reasons by removing all name and URL information from the database, only leaving a strip-down version linking EJO ID and Conflict ID.

\subsection*{Data Analysis}
\begin{figure}[ht!]
\centering
\includegraphics[width=1.0\textwidth]{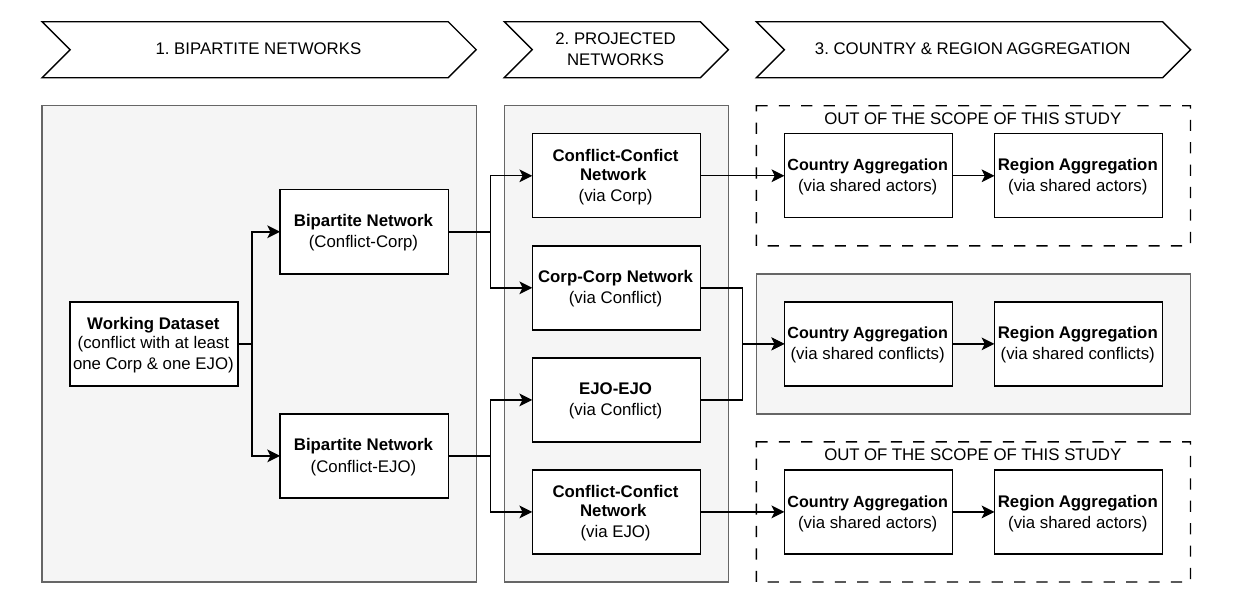}
\caption{\textbf{Flowchart representing the three-step methodology.} Starting from the bipartite (actor-conflict) networks, these have been projected (via conflict or via actor) and finally aggregated at country or regional level. The country and regional aggregation of the conflict-conflict networks remain out of the scope of this study. }
\label{fig 1: flowchart}
\end{figure}

Figure \ref{fig 1: flowchart} shows the methodological flowchart highlighting the three-step methodology for this study. First, two bipartite networks (Conflict--EJO AND conflict--company) have been built. Secondly, from these two bipartite networks and through the co-occurrence of actors in environmental conflicts, two projected networks have been obtained, one for EJOs and one for companies. Thirdly and finally, the projected networks have been aggregated both at the country and regional level (via shared conflicts) by considering the country of the legal/administrative headquarters for each actor. 

\begin{figure}[ht!]
\centering
\includegraphics[width=1.0\textwidth]{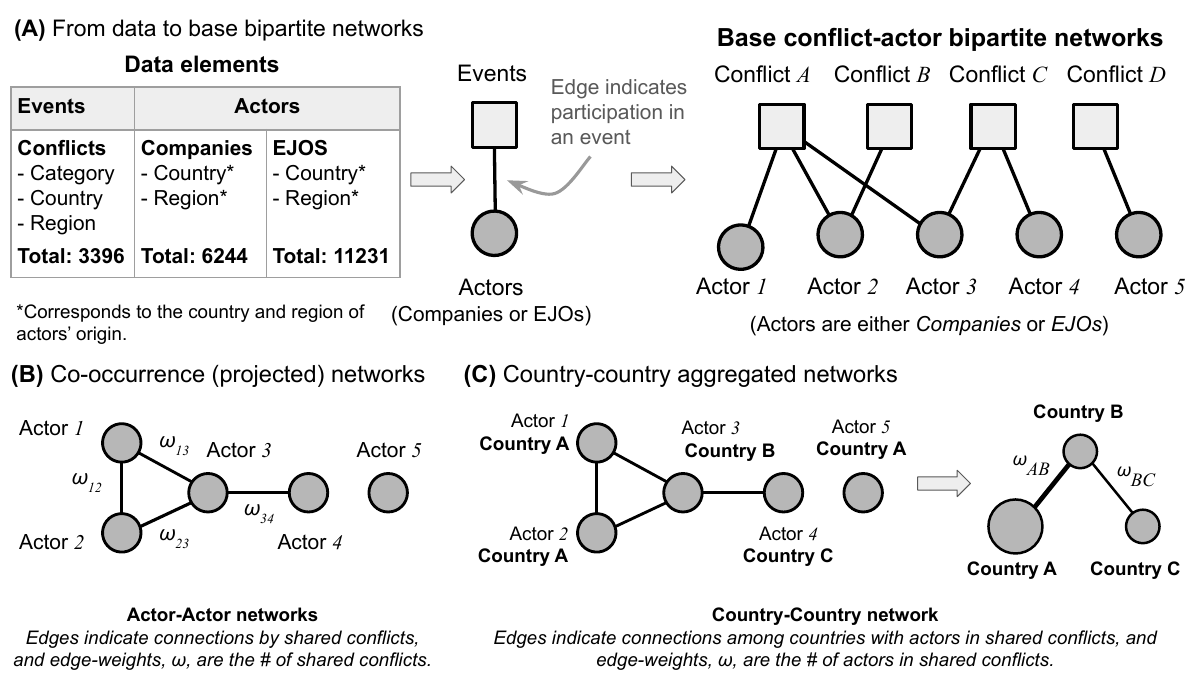}
\caption{\textit{Visual representation of the three-step methodology.} (A) Working dataset and bipartite, conflict--actor, network representation. Squares represent conflicts while circles represent actors, either companies or EJOs. For each node, either a conflict or an actor, only the category, country and region information have been used. In the bipartite network, edges represent the involvement of an actor in a conflict. (B) Projected actor networks. In the monolayer actor--actor network, an edge represent the co-occurrence of two actors, either EJOs or companies. The edge weight measure the number of shared conflict between two actors. Similarly, a conflict--conflict monolayer projected network can be obtained by looking at conflicts connected through a shared EJO or company, i.e. an EJO/company involved in two conflicts. (C) country--country (or region--region) aggregated networks. Organizations or conflicts can be assigned and aggregated per country or region. An edge between two country represent if the two country have connections between two, or more, companies headquartered in those countries.}
\label{fig 2: visual representation}
\end{figure}

In the following paragraphs, all details regarding the bipartite and projected networks, as well as the country--country aggregated networks will be described. Figure \ref{fig 2: visual representation} shows graphically and formally the three-step methodology.

\paragraph{(A) From data to base bipartite networks.} Fig. \ref{fig 2: visual representation}A shows the procedure from data to the base bipartite networks: the conflict-actor networks. The underlying hypothesis used to build bipartite networks is the involvement (with no formal distinction on the type of participation) of an actor within a specific conflict. The EJAtlas, among other useful information \citep{temper2015mapping}, can be conceptualized as composed by three main \textit{data elements}, \textit{conflicts},\textit{companies}, and \textit{EJOs}, each one with certain information. Each \textit{conflict} has a \textit{category}, a \textit{country} of occurrence and consequently a \textit{region}.
The full list of categories in EJAtlas is: Mining, Fossil Fuels and Energy, Biomass and Land, Water Management, Infrastructures, Industrial and Utilities, Waste Management, Tourism Recreation, Nuclear, and Biodiversity Conservation. The considered regions in this study are sub-continental areas: Europe, Africa, North America, Latin America, Pacific, China, Russia and Central Asia, South Asia, South East Asia, Middle East, East Asia. Each \textit{company} has a \textit{country} and \textit{region} assigned based on the location of their headquarters. \textit{EJOs} do not have a \textit{country} assigned in the database. In this study, we assign a \textit{country} to an EJO when more than 50\% of the \textit{conflicts} it is involved in occur in a given \textit{country}. Whenever an EJO has not at least 50\% of the conflicts from a single specific country, they are considered \textit{International} organizations. This labelling rule captures transnational environmental justice organizations that operate across multiple countries and therefore cannot be meaningfully associated with a single national context.

Formally, let $A=\{a_1,...,a_n\}$ denote the set of EJOs and $C=\{c_1,...,c_m\}$ the set of conflicts. For each EJO $a_i$, we define the set of conflicts in which the EJO participates as $C(a_i) = \{c_j \in C \mid a_i \text{ participates in } c_j\}$. Let $G(a_i,c_j)$ denote the country associated with EJO $a_i$ in conflict $c_j$. Therefore, each conflict $c_j$ is associated with a specific country, denoted by $g(c_j)$. For each EJO $a_i$, we consider the set of conflicts in which the organization participates, defined as $C(a_i)$. We then compute the number of conflicts associated with each country: $N_g(a_i) = \sum_{c_j \in C(a_i)} \mathbf{1}\big(g(c_j)=g\big)$, where $\mathbf{1}(\cdot)$ denotes the indicator function. A country $g^*$ is assigned to the EJO if and only if the majority of conflicts involving that EJO occur within the same country, that is $g^* = \arg\max_g N_g(a_i) \quad \text{if} \quad \frac{N_{g^*}(a_i)}{|C(a_i)|} \ge 0.5.$. If no country accounts for at least $50\%$ of the conflicts involving the organization, the EJO is classified as \textit{International}. Formally, the country assignment rule can be written as:
\[
\hat{g}(a_i) =
\begin{cases}
g^*, & \text{if } \frac{N_{g^*}(a_i)}{|C(a_i)|} \ge 0.5 \\
\text{International}, & \text{otherwise}.
\end{cases}
\].

Regarding the company and EJO \textit{category}, since actors may appear in multiple environmental conflicts with different categories, each actor may belong to several categories based on the conflict category to which it is related with. Formally, let define $L(c_j)$ denote the category associated with conflict $c_j$. The set of categories assigned to actor $a_i$ is then defined as $\mathcal{L}(a_i) = \{\, L(c_j) \mid c_j \in C(a_i) \,\}$.

\paragraph{(B) Co-occurrence (projected) networks.} Fig.~\ref{fig 2: visual representation}B shows the rationale behind the projected networks.
The underlying hypothesis, in this step, is the co-occurrence in a conflict of an organization. In other words, an edge exists if two actors (either an EJO or a company) have at least one shared conflict, while the weight $\omega_ij$ is the $\#$ of shared conflicts among actor $i$ and actor $j$. It is noteworthy to specify that co-occurrence does not imply any legal or formal collaborations between two actors but simply, it represents their simultaneous involvement in an environmental conflict. The specific details, as well as any legal implications, must be analyzed on a case-by-case basis.

Formally, let $B \in \{0,1\}^{m \times n}$ be the bipartite adjacency matrix linking conflicts (rows) to actors (columns), where:
\[
B_{k i} =
\begin{cases}
1 & \text{if actor } i \text{ is involved in conflict } k \\
0 & \text{otherwise}
\end{cases}
\]
The weight of the edge between two nodes $i$ and $j$ in the projected network is defined as: $w_{ij} = \sum_{k=1}^{m} B_{k i} \cdot B_{k j}, \quad i \neq j $ which corresponds to the number of conflicts in which both nodes co-occur. In matrix form, the weighted projection is obtained as: $ W = B^\top B $.

Two types of projected networks are presented: (1) projections \textit{via conflict}, i.e., company/EJO networks, and (2) projections \textit{via company/EJO}, i.e., conflict networks.

First (Fig.~\ref{fig 5: actor-actor network}), each projected actor network, then, has been analyzed through different centrality degrees: i) average weight (the $\#$ of conflicts shared between two actors), ii) node degree (the $\#$ of actor a given actor is linked to), iii) betweenness (i.e., the property of a node to act as a bridge in the global network, or in other words the measure of the \# of shortest path between all other pairs of nodes in a graph passing through a specific node) and iv) closeness (a measure of how a node is close to all other nodes in a graph by computing the inverse of the sum of its shortest path distances to everyone else) centralities \cite{zhang2017degree}. For both network min, mean, and max values have been computed, as well as the Gini coefficient for each indicator. Formally, for a set of values $x = \{x_1, x_2, \dots, x_n\}$ sorted in non-decreasing order ($x_i \leq x_{i+1}$), the Gini coefficient is defined as $G = \frac{1}{2 n^2 \mu} \sum_{i=1}^{n} \sum_{j=1}^{n} |x_i - x_j| $ where $n$ is the number of observations and $\mu = \frac{1}{n}\sum_{i=1}^{n} x_i$ is the mean of the distribution. The Gini coefficient takes values in the interval $[0,1]$, where $G=0$ indicates perfect equality (all nodes have identical values) and $G \to 1$ indicates maximum inequality (a small number of nodes concentrate most of the value). In the context of network analysis, higher Gini values reveal stronger concentration of connectivity or centrality in a few nodes (i.e., the presence of hubs or highly central actors), while lower values indicate a more homogeneous and evenly distributed network structure. Moreover, for both networks several other information and global indicators have been evaluated to compare the two networks. Specifically, we computed the density, clustering, mean degree and path length, diameter, number of connected components and the size (in percentage over the total of nodes) of the largest connected components. Furthermore, a stress test have been conducted to evaluate the robustness of the global networks, both for EJOs and companies, by removing the first top nodes with higher degree. Specifically, this stress test was performed to evaluate the resistance of LCCs to high-degree nodes. The LCC's size of the reduced networks have been computed to quantitatively evaluate how the entire network depends on top-degree actors. 

Second (Fig.~\ref{fig 6: conflict-conflict network}), each projected conflict network (i.e., projected via companies or via EJOs) has been analyzed and compared via density, clustering, mean degree and path length, diameter, number of connected components and the size of the LCC  similarly to what was done for the actor networks. Moreover, since conflicts have a pre-assigned category, the edge frequency by category - both intra and inter category - i.e. within same category or between two different categories - have been analyzed as a measure of heterogeneity / homogeneity of the network. A higher intra-category edge frequency represents the case when actors co-occurred mostly in similar conflicts (Fossil, Mineral, ...), while a high inter-category edge frequency shows if actors co-occurred in several conflicts independently by a specific category.

\paragraph{(C) Country--country aggregated networks.} 
In terms of geographical information, aggregation itself may be conducted according to different logic aiming at highlighting different effects. Specifically, each country node may represent the $\#$ of actors or conflicts in the country, while the links may map both the $\#$ of actor--actor edges via shared conflicts (as described in Fig.~\ref{fig 2: visual representation}C) or the number of country--country links, i.e., the $\#$ of countries connected with a specific country via company relationships. In the former case (out of the scope of this study), information highlight mainly global distribution of actors (i.e., companies mainly situated in the global North and EJOs in the global South - information partially discussed in \citet{llavero2025driving}) - while in the latter the aggregated visualization highlights structural dependencies among countries. It is necessary to point out that in both these visualizations information regarding conflict locations is lost. 

Fig.~\ref{fig 2: visual representation}C represents aggregations at country (or region) level. In this case, the starting point are the company and the EJO networks, where two actors are connected if they co-occur in at least one shared conflict. The country--country network is then obtained by aggregating actors according to their assigned country. Therefore, each node in the aggregated network represents a country, and its size is proportional to the number of actors assigned to that country. Formally, let $A=\{a_1,\dots,a_n\}$ be the set of actors and let $g(a_i)$ denote the country assigned to actor $a_i$. Let $w_{ij}$ denote the weight of the edge between actors $a_i$ and $a_j$ in the projected actor--actor network, where $w_{ij}$ corresponds to the number of shared conflicts in which the two actors co-occur. The weight between two countries $p$ and $q$ in the aggregated country--country network is then defined as the sum of all co-occurrences between actors headquartered in those two countries:
\[
w_{pq} = \sum_{i=1}^{n}\sum_{j=1}^{n} w_{ij}\,\mathbf{1}\big(g(a_i)=p\big)\,\mathbf{1}\big(g(a_j)=q\big),
\qquad p \neq q
\]
where $\mathbf{1}(\cdot)$ is the indicator function. Thus, $w_{pq}$ measures the total number of co-occurrences in shared conflicts between actors belonging to country $p$ and actors belonging to country $q$. Similarly, the size of each country node is defined as the number of actors headquartered in that country:

\[
s_p = \sum_{i=1}^{n} \mathbf{1}\big(g(a_i)=p\big)
\].

As represented in Fig.~\ref{fig 2: visual representation}C, let us suppose that \textit{Actor 1} and \textit{Actor 2} belong to \textit{Country A}, while \textit{Actor 3} belongs to \textit{Country B} and \textit{Actor 4} belongs to \textit{Country C}. If \textit{Actor 1} and \textit{Actor 2} are both connected to \textit{Actor 3}, and \textit{Actor 3} is connected to \textit{Actor 4}, then the aggregated country--country network will contain an edge between \textit{Country A} and \textit{Country B} with weight $w_{AB}$ equal to the sum of co-occurrences between actors in those two countries, and an edge between \textit{Country B} and \textit{Country C} with weight $w_{BC}$ defined analogously.

Similarly, we analyzed the \textit{country degree}, defined as the number of distinct countries connected to a given country. Formally, let $C$ be the set of countries and let $w_{pq}$ denote the weight of the edge between country $p$ and country $q$, defined as the total number of co-occurrences between actors headquartered in $p$ and $q$. The country degree of country $p$ is defined as:

\[
k_p = \sum_{q \in C, \, q \neq p} \mathbf{1}(w_{pq} > 0)
\]

where $\mathbf{1}(\cdot)$ is the indicator function. Thus, $k_p$ counts the number of distinct countries with which country $p$ shares at least one co-occurrence in environmental conflicts.

Therefore, in this final step, the information about the country where a conflict takes place is no longer retained explicitly, since aggregation is performed on the basis of actors' headquarters. In addition, any information about intra-country links is lost. However, this transformation makes it possible to investigate the transnational relations to identify which national contexts are more interconnected through actors participating in shared environmental conflicts.

\pagebreak
{\small
\bigskip
\section*{Data availability} Data is available from Marcel Llavero-Pasquina and the EJAtlas project upon reasonable request. 

\medskip
\section*{Conflict of interests} The research was conducted in the absence of any commercial or financial relationships that could be construed as a potential conflict of interest.

\medskip
\section*{Authors contributions} DC: Conceptualization, Data Analysis \& Visualization, Writing \& Review, Methodology. JRNC: Conceptualization, Data Analysis, Writing \& Review, Methodology. MLP: Conceptualization, Data Analysis, Writing \& Review, Methodology. JMA:  Writing \& Review.

\medskip
\section*{Acknowledgements} JRNC acknowledges support from Mexico's Ministry of Science, Humanities, Technology and Innovation (SECIHTI), under the program ``Estancias Posdoctorales por México 2023''. MLP acknowledges support from the Holberg price awarded to JMA, and from the Maria de Maeztu grant from the Spanish Ministry of Science (CEX2024-001506-M) awarded to ICTA-UAB.
}


{\footnotesize
\bibliographystyle{apalike}
\bibliography{references}

@incollection{della2021transnational,
  title={Transnational activisms and the global justice movement},
  author={Della Porta, Donatella and Marchetti, Raffaele},
  booktitle={Routledge International Handbook of Contemporary Social and Political Theory},
  pages={466--477},
  year={2021},
  publisher={Routledge}
}

@article{smith2013transnational,
  title={Transnational social movements},
  author={Smith, Jackie},
  journal={The Wiley-Blackwell Encyclopedia of Social and Political Movements},
  year={2013},
  publisher={Wiley Online Library}
}

@article{almeida2018globalization,
  title={Globalization and social movements},
  author={Almeida, Paul and Chase-Dunn, Chris},
  journal={Annual Review of Sociology},
  volume={44},
  number={1},
  pages={189--211},
  year={2018},
  publisher={Annual Reviews}
}

@article{llavero2024political,
  title={The political ecology of oil and gas corporations: TotalEnergies and post-colonial exploitation to concentrate energy in industrial economies},
  author={Llavero-Pasquina, Marcel and Navas, Grettel and Cantoni, Roberto and Mart{\'\i}nez-Alier, Joan},
  journal={Energy Research \& Social Science},
  volume={109},
  pages={103434},
  year={2024},
  publisher={Elsevier}
}

@book{fressoz2024sans,
  title={Sans transition: une nouvelle histoire de l'{\'e}nergie},
  author={Fressoz, Jean-Baptiste},
  year={2024},
  publisher={Seuil}
}

@article{Martinez-Alier03052016,
author = {Joan Martinez-Alier and Leah Temper and Daniela Del Bene and Arnim Scheidel},
title = {Is there a global environmental justice movement?},
journal = {The Journal of Peasant Studies},
volume = {43},
number = {3},
pages = {731--755},
year = {2016},
publisher = {Routledge},
doi = {10.1080/03066150.2016.1141198},
URL = {https://doi.org/10.1080/03066150.2016.1141198},
eprint = {https://doi.org/10.1080/03066150.2016.1141198}
}

@article{menton2020environmental,
  title={Environmental justice and the SDGs: from synergies to gaps and contradictions},
  author={Menton, Mary and Larrea, Carlos and Latorre, Sara and Martinez-Alier, Joan and Peck, Mika and Temper, Leah and Walter, Mariana},
  journal={Sustainability science},
  volume={15},
  number={6},
  pages={1621--1636},
  year={2020},
  publisher={Springer}
}

@article{temper2018perspective,
  title={A perspective on radical transformations to sustainability: resistances, movements and alternatives},
  author={Temper, Leah and Walter, Mariana and Rodriguez, Ioki{\~n}e and Kothari, Ashish and Turhan, Ethemcan},
  journal={Sustainability Science},
  volume={13},
  number={3},
  pages={747--764},
  year={2018},
  publisher={Springer}
}

@article{hickel2022imperialist,
  title={Imperialist appropriation in the world economy: Drain from the global South through unequal exchange, 1990--2015},
  author={Hickel, Jason and Dorninger, Christian and Wieland, Hanspeter and Suwandi, Intan},
  journal={Global environmental change},
  volume={73},
  pages={102467},
  year={2022},
  publisher={Elsevier}
}

@article{llavero2025driving,
  title={Driving ecologically unequal exchange: A global analysis of multinational corporations’ role in environmental conflicts},
  author={Llavero-Pasquina, Marcel},
  journal={Global Environmental Change},
  volume={92},
  pages={103006},
  year={2025},
  publisher={Elsevier}
}

@article{cantoni2025franccafrique,
  title={From Fran{\c{c}}afrique to Chinafrica? Ecologically unequal exchange, neocolonialism, and environmental conflicts in Africa},
  author={Cantoni, Roberto and Llavero-Pasquina, Marcel and Apostolopoulou, Elia and Gerber, Julien-Fran{\c{c}}ois and Bond, Patrick and Martinez-Alier, Joan},
  journal={World Development},
  volume={193},
  pages={107015},
  year={2025},
  publisher={Elsevier}
}

@article{scheidel2020environmental,
  title={Environmental conflicts and defenders: A global overview},
  author={Scheidel, Arnim and Del Bene, Daniela and Liu, Juan and Navas, Grettel and Mingorr{\'\i}a, Sara and Demaria, Federico and Avila, Sof{\'\i}a and Roy, Brototi and Ert{\"o}r, Irmak and Temper, Leah and others},
  journal={Global Environmental Change},
  volume={63},
  pages={102104},
  year={2020},
  publisher={Elsevier}
}

@article{atapattu2018extractive,
  title={Extractive industries and inequality: Intersections of environmental law, human rights, and environmental justice},
  author={Atapattu, Sumudu},
  journal={Ariz. St. LJ},
  volume={50},
  pages={431},
  year={2018},
  publisher={HeinOnline}
}

@article{dowling2025mirages,
  title={Mirages or miracles? Lithium extraction and the just energy transition},
  author={Dowling, Carmel and Otero, Gerardo},
  journal={Energy Research \& Social Science},
  volume={119},
  pages={103862},
  year={2025},
  publisher={Elsevier}
}

@techreport{IEA2021,
  author = {IEA},
  year = {2021},
  title = {The Role of Critical Minerals in Clean Energy Transitions, IEA},
  address = {Paris},
  institution = {International Energy Agency},
  note = {Licence: CC BY 4.0},
  url = {https://www.iea.org/reports/the-role-of-critical-minerals-in-clean-energy-transitions},
}

@article{stammler2016resources,
  title={Resources, rights and communities: Extractive mega-projects and local people in the Russian Arctic},
  author={Stammler, Florian and Ivanova, Aitalina},
  journal={Europe-Asia Studies},
  volume={68},
  number={7},
  pages={1220--1244},
  year={2016},
  publisher={Taylor \& Francis}
}

@article{orihuela2025megaprojects,
  title={Megaprojects Observed: The Extractivist Hiding Hand at Resource Peripheries},
  author={Orihuela, Jos{\'e} Carlos and Serrano, Sergio},
  journal={The Journal of Environment \& Development},
  volume={34},
  number={4},
  pages={872--894},
  year={2025},
  publisher={SAGE Publications Sage CA: Los Angeles, CA}
}

@article{busscher2020environmental,
  title={Environmental justice implications of land grabbing for industrial agriculture and forestry in Argentina},
  author={Busscher, Nienke and Parra, Constanza and Vanclay, Frank},
  journal={Journal of Environmental Planning and Management},
  volume={63},
  number={3},
  pages={500--522},
  year={2020},
  publisher={Taylor \& Francis}
}

@article{temper2015mapping,
  title={Mapping the frontiers and front lines of global environmental justice: the EJAtlas},
  author={Temper, Leah and Del Bene, Daniela and Martinez-Alier, Joan},
  journal={Journal of political ecology},
  volume={22},
  number={1},
  year={2015},
  publisher={University of Arizona Libraries}
}

@incollection{cantoni2025ejatlas,
  title={The EJAtlas: co-production of knowledge for engaged research to support environmental justice movements, education, and policy-making},
  author={Cantoni, Roberto and Del Bene, Daniela and Fanari, Eleonora and Llavero-Pasquina, Marcel and Scheidel, Arnim and Walter, Mariana},
  booktitle={The new Routledge handbook of political ecology},
  pages={230--237},
  publisher={Routledge},
  year={2025}
}

@article{martinez2021mapping,
  title={Mapping ecological distribution conflicts: The EJAtlas},
  author={Martinez-Alier, Joan},
  journal={The Extractive Industries and Society},
  volume={8},
  number={4},
  pages={100883},
  year={2021},
  publisher={Elsevier}
}

@article{Carroll2003,
author = {Carroll, William K. and Carson, Colin},
title = {The network of global corporations and elite policy groups: a structure for transnational capitalist class formation?},
journal = {Global Networks},
volume = {3},
number = {1},
pages = {29-57},
doi = {https://doi.org/10.1111/1471-0374.00049},
url = {https://onlinelibrary.wiley.com/doi/abs/10.1111/1471-0374.00049},
eprint = {https://onlinelibrary.wiley.com/doi/pdf/10.1111/1471-0374.00049},
year = {2003}
}

@article{aydin2017network,
  title={Network effects in environmental justice struggles: An investigation of conflicts between mining companies and civil society organizations from a network perspective},
  author={Aydin, Cem Iskender and Ozkaynak, Begum and Rodr{\'\i}guez-Labajos, Beatriz and Yenilmez, Taylan},
  journal={PloS one},
  volume={12},
  number={7},
  pages={e0180494},
  year={2017},
  publisher={Public Library of Science San Francisco, CA USA}
}

@article{vitali2014community,
  title={The community structure of the global corporate network},
  author={Vitali, Stefania and Battiston, Stefano},
  journal={PloS one},
  volume={9},
  number={8},
  pages={e104655},
  year={2014},
  publisher={Public Library of Science San Francisco, USA}
}

@misc{de2019complexity,
  title={Complexity explained},
  author={De Domenico, Manlio and Brockmann, Dirk and Camargo, CQ and Gershenson, Carlos and Goldsmith, Daniel and Jeschonnek, Sabine and Kay, Lorren and Nichele, Stefano and Nicol{\'a}s, J and Schmickl, Thomas and others},
  year={2019},
  publisher={University of Exeter}
}

@article{de2022multilayer,
  title={Multilayer networks: analysis and visualization},
  author={De Domenico, Manlio},
  journal={Introduction to muxViz with R. Cham: Springer},
  year={2022},
  publisher={Springer}
}

@article{ladyman2013complex,
  title={What is a complex system?},
  author={Ladyman, James and Lambert, James and Wiesner, Karoline},
  journal={European Journal for Philosophy of Science},
  volume={3},
  number={1},
  pages={33--67},
  year={2013},
  publisher={Springer}
}

@article{aleta2025multilayer,
  title={Multilayer network science: theory, methods, and applications},
  author={Aleta, Alberto and Teixeira, Andreia Sofia and de Arruda, Guilherme Ferraz and Baronchelli, Andrea and Barrat, Alain and Kert{\'e}sz, J{\'a}nos and D{\'\i}az-Guilera, Albert and Artime, Oriol and Starnini, Michele and Petri, Giovanni and others},
  journal={arXiv preprint arXiv:2511.23371},
  year={2025}
}

@article{luna2020corruption,
  title={Corruption and complexity: a scientific framework for the analysis of corruption networks},
  author={Luna-Pla, Issa and Nicol{\'a}s-Carlock, Jos{\'e} R},
  journal={Applied Network Science},
  volume={5},
  number={1},
  pages={13},
  year={2020},
  publisher={Springer}
}

@article{llavero2025false,
  title={False solutions: How do fossil fuel companies reproduce their power through the energy transition?},
  author={Llavero-Pasquina, Marcel and Eckstein, Sarah and Daley, Freddie and Rugh, Nathaniel and Glimmerveen, Judith},
  journal={Energy Research \& Social Science},
  pages={104367},
  year={2025},
  publisher={Elsevier}
}

@book{barabasi2011bursts,
  title={Bursts: the hidden patterns behind everything we do, from your e-mail to bloody crusades},
  author={Barab{\'a}si, Albert-L{\'a}szl{\'o}},
  year={2011},
  publisher={Penguin}
}

@article{castellano2009statistical,
  title={Statistical physics of social dynamics},
  author={Castellano, Claudio and Fortunato, Santo and Loreto, Vittorio},
  journal={Reviews of modern physics},
  volume={81},
  number={2},
  pages={591--646},
  year={2009},
  publisher={APS}
}

@article{Newman2006,
  author = {Newman, Mark E. J.},
  title = {Modularity and community structure in networks},
  journal = {Proceedings of the National Academy of Sciences},
  volume = {103},
  number = {23},
  pages = {8577--8582},
  year = {2006},
  doi = {10.1073/pnas.0601602103}
}

@article{fortunato2010community,
  title={Community detection in graphs},
  author={Fortunato, Santo},
  journal={Physics reports},
  volume={486},
  number={3-5},
  pages={75--174},
  year={2010},
  publisher={Elsevier}
}

@inproceedings{zhang2017degree,
  title={Degree centrality, betweenness centrality, and closeness centrality in social network},
  author={Zhang, Junlong and Luo, Yu},
  booktitle={2017 2nd international conference on modelling, simulation and applied mathematics (MSAM2017)},
  pages={300--303},
  year={2017},
  organization={Atlantis press}
}

@article{blondel2008fast,
  title={Fast unfolding of communities in large networks},
  author={Blondel, Vincent D and Guillaume, Jean-Loup and Lambiotte, Renaud and Lefebvre, Etienne},
  journal={Journal of statistical mechanics: theory and experiment},
  volume={2008},
  number={10},
  pages={P10008},
  year={2008}
}

@article{helbing2013globally,
  title={Globally networked risks and how to respond},
  author={Helbing, Dirk},
  journal={Nature},
  volume={497},
  number={7447},
  pages={51--59},
  year={2013},
  publisher={Nature Publishing Group UK London}
}

@book{ghosh2019dependency,
  title={Dependency theory revisited},
  author={Ghosh, Baidyanath N},
  year={2019},
  publisher={Routledge}
}

@article{lindell2009glocal,
  title={‘Glocal’movements: place struggles and transnational organizing by informal workers},
  author={Lindell, Ilda},
  journal={Geografiska Annaler: Series B, Human Geography},
  volume={91},
  number={2},
  pages={123--136},
  year={2009},
  publisher={Taylor \& Francis}
}

@article{vanhaute2014globalizing,
  title={Globalizing local struggles. Localizing global struggles. Peasant movements from local to global platforms and back},
  author={Vanhaute, Eric},
  journal={Workers of the World. International Journal on Strikes and Social Conflict},
  volume={1},
  number={5},
  pages={114--129},
  year={2014}
}

@Inbook{Carlock2021,
author="Nicol{\'a}s-Carlock, Jos{\'e} R. and Luna-Pla, Issa",
title="Corruptomics",
bookTitle="Corruption Networks: Concepts and Applications",
year="2021",
publisher="Springer International Publishing",
address="Cham",
pages="153--158",
isbn="978-3-030-81484-7",
doi="10.1007/978-3-030-81484-7_9",
url="https://doi.org/10.1007/978-3-030-81484-7_9"
}

@article{nicolas2021conspiracy,
  title={Conspiracy of corporate networks in corruption scandals},
  author={Nicol{\'a}s-Carlock, Jos{\'e} R and Luna-Pla, Issa},
  journal={Frontiers in Physics},
  volume={9},
  pages={667471},
  year={2021},
  publisher={Frontiers Media SA}
}

@incollection{mliless2022reporting,
  title={Reporting International Conflicts Through the Environmental Discourse: The Moroccan Sahara Conflict as a Case Study},
  author={Mliless, Mohamed and Larouz, Mohammed},
  booktitle={The Climate-Conflict-Displacement Nexus from a Human Security Perspective},
  pages={373--403},
  year={2022},
  publisher={Springer}
}

@phdthesis{christiansen2025corruption,
  title={Corruption and Foreign Influence: A Cross-National Analysis of Environmental Violence},
  author={Christiansen, Jamilah D},
  year={2025},
  school={University of South Florida}
}

@incollection{williams2024transnational,
  title={Transnational environmental crime and corruption},
  author={Williams, David Aled},
  booktitle={Research handbook on environmental crimes and criminal enforcement},
  pages={246--262},
  year={2024},
  publisher={Edward Elgar Publishing}
}

@article{avila2018environmental,
  title={Environmental justice and the expanding geography of wind power conflicts},
  author={Avila, Sofia},
  journal={Sustainability Science},
  volume={13},
  number={3},
  pages={599--616},
  year={2018},
  publisher={Springer}
}

@article{sebastien2019resistance,
  title={Resistance as an enlightening process: a new framework for analysis of the socio-political impacts of place-based environmental struggles},
  author={Sebastien, Lea and Pelenc, Jerome and Milanesi, Julien},
  journal={Local Environment},
  volume={24},
  number={5},
  pages={487--504},
  year={2019},
  publisher={Taylor \& Francis}
}

@article{Kimberly2012,
author = {Creasap, Kimberly},
title = {Social Movement Scenes: Place-Based Politics and Everyday Resistance},
journal = {Sociology Compass},
volume = {6},
number = {2},
pages = {182-191},
doi = {https://doi.org/10.1111/j.1751-9020.2011.00441.x},
url = {https://compass.onlinelibrary.wiley.com/doi/abs/10.1111/j.1751-9020.2011.00441.x},
year = {2012}
}

@article{martinezzijin,
  title={Zijin: A Growing Metal Mining Chinese Transnational Firm},
  author={Martinez-Alier, Joan and Llavero-Pasquina, Marcel},
  journal={Commodity Frontiers},
  volume={7},
  pages={63--106},
  year={2026}
}

@article{neyra2026glencore,
  author  = {Neyra, Raquel and Martinez-Alier, Joan},
  title   = {Glencore, a Giant Facing Its Contradictions: Social and Environmental Unaccounted Costs},
  journal = {Extractive Industries and Society},
  year    = {2026},
  note    = {Accepted for publication. Manuscript no. EXIS-D-25-00744}
}

@book{demaria2025contested,
  editor    = {Demaria, Federico and Vico, Daniela and Gabard, Lucie F.},
  title     = {Contested Waste: Environmental Conflicts and Waste Picker Resistance in the Global South},
  edition   = {1},
  year      = {2025},
  publisher = {Routledge},
  doi       = {10.4324/9781003468516}
}

@incollection{georgescu-roegen1971entropy,
  author       = {Georgescu-Roegen, Nicholas},
  title        = {The Entropy Law and the Economic Problem},
  booktitle    = {Valuing the Earth: Economics, Ecology, Ethics},
  editor       = {Daly, Herman E. and Townsend, Kenneth N.},
  publisher    = {MIT Press},
  address      = {Cambridge, MA},
  year         = {1971},
  origdate     = {1971},
  pages        = {75--88},
  isbn         = {9780262260565}
}

@book{martinez2023land,
  title={Land, water, air and freedom: The making of world movements for environmental justice},
  author={Mart{\'\i}nez-Alier, Joan},
  year={2023},
  publisher={Edward Elgar Publishing}
}

@article{Temper_2020,
doi = {10.1088/1748-9326/abc197},
year = {2020},
publisher = {IOP Publishing},
volume = {15},
number = {12},
pages = {123004},
author = {Temper, Leah and Avila, Sofia and Bene, Daniela Del and Gobby, Jennifer and Kosoy, Nicolas and Billon, Philippe Le and Martinez-Alier, Joan and Perkins, Patricia and Roy, Brototi and Scheidel, Arnim and Walter, Mariana},
title = {Movements shaping climate futures: A systematic mapping of protests against fossil fuel and low-carbon energy projects},
journal = {Environmental Research Letters}
}

@article{FischerKowalski01011997,
author = {Marina Fischer‐Kowalski and Helmut Haberl},
title = {Tons, joules, and money: Modes of production and their sustainability problems},
journal = {Society \& Natural Resources},
volume = {10},
number = {1},
pages = {61--85},
year = {1997},
publisher = {Routledge},
doi = {10.1080/08941929709381009}
}

@article{hanavcek2024we,
  title={“We are protectors, not protestors”: global impacts of extractivism on human--nature bonds},
  author={Hana{\v{c}}ek, Ksenija and Tran, Dalena and Landau, Arielle and Sanz, Teresa and Thiri, May Aye and Navas, Grettel and Del Bene, Daniela and Liu, Juan and Walter, Mariana and Lopez, Aida and others},
  journal={Sustainability science},
  volume={19},
  number={6},
  pages={1789--1808},
  year={2024},
  publisher={Springer}
}

@book{bunker1985underdeveloping,
  author    = {Bunker, Stephen G.},
  title     = {Underdeveloping the Amazon: Extraction, Unequal Exchange, and the Failure of the Modern State},
  year      = {1985},
  publisher = {University of Illinois Press},
}

@article{Martinez-Alier02112022,
author = {Joan Martinez-Alier},
title = {Circularity, entropy, ecological conflicts and LFFU},
journal = {Local Environment},
volume = {27},
number = {10-11},
pages = {1182--1207},
year = {2022},
publisher = {Routledge},
doi = {10.1080/13549839.2021.1983795},
}

@incollection{Bosma2024,
    author = {Bosma, Ulbe and Vanhaute, Eric},
    editor = {Curry-Machado, Jonathan and Stubbs, Jean and Clarence-Smith, William Gervase and Vos, Jelmer},
    isbn = {9780197502679},
    title = {Commodity Frontiers: Linking Global Capitalism and Local Resilience},
    booktitle = {The Oxford Handbook of Commodity History},
    publisher = {Oxford University Press},
    year = {2024},
    doi = {10.1093/oxfordhb/9780197502679.013.5}
}

@book{polimeni2015myth,
  title={The myth of resource efficiency: The Jevons paradox},
  author={Polimeni, John M and Mayumi, Kozo and Giampietro, Mario and Alcott, Blake},
  year={2015},
  publisher={Routledge}
}

@article{d2024accumulation,
  title={Accumulation by contamination: Worldwide cost-shifting strategies of capital in waste management},
  author={D’Alisa, Giacomo and Demaria, Federico},
  journal={World development},
  volume={184},
  pages={106725},
  year={2024},
  publisher={Elsevier}
}
}


\end{document}